\documentclass[useAMS,usenatbib]
{mn2e}
\usepackage{epsfig}
\usepackage{amsmath, amssymb,bm}

\def \be{\begin{equation}}
\def \ee{\end{equation}}
\def \rsun{\rm R_{\odot}}
\def \msun{\rm M_{\odot}}

\begin{document}
\title[]{ One for the Future: measuring the mass transfer rate in the ULX M82 X-2 by using orbital period changes will take millenia
}

\author[Andrew King \& Jean--Pierre Lasota] 
{\parbox{5in}{Andrew King$^{1, 2, 3}$ \& Jean--Pierre Lasota$^{4, 5}$
}
\vspace{0.1in} \\ $^1$ School of Physics \& Astronomy, University
of Leicester, Leicester LE1 7RH UK\\ 
$^2$ Astronomical Institute Anton Pannekoek, University of Amsterdam, Science Park 904, NL-1098 XH Amsterdam, The Netherlands \\
$^{3}$ Leiden Observatory, Leiden University, Niels Bohrweg 2, NL-2333 CA Leiden, The Netherlands\\
$^{4}$ Institut d'Astrophysique de Paris, CNRS et Sorbonne Universit\'e, UMR 7095, 98bis Bd Arago, 75014 Paris, France\\  
$^{5}$ Nicolaus Copernicus Astronomical Center, Polish Academy of Sciences, ul. Bartycka 18, 00-716 Warsaw, Poland\\     
}


\maketitle

\begin{abstract}
\citet{Bachettietal21} have recently claimed to measure the mass transfer rate in the pulsing ULX system M82 X-2 
by following the change of its orbital period over 7 yr. We reiterate the known point that this method cannot give a reliable result (or even necessarily predict the correct sign of long--term period change)
without a far longer baseline (here $\gg 1000$~yr)
or for systems with a much higher long--term mass transfer rate ($\gg 10^{-4}\msun\,{\rm yr}^{-1}$), if they exist. Applying the method of \citet{Bachettietal21}
to measured orbital period derivatives predicts that the 
well--studied quiescent X--ray transients XTEJ1118+480, A0620--00 should currently instead have steady accretion discs and be
bright X--ray sources, while Nova Muscae 1991 should be still brighter (a ULX). But all three sources are observed to be extremely faint.
We conclude that there is no evidence to support the high mass transfer rate that \citet{Bachettietal21} find for M82 X-2 as it is deduced from period noise not related to the binary evolution.
\end{abstract}

\footnotetext[1]{E-mail: ark@astro.le.ac.uk}
\section{Introduction}
\label{intro}
\citet{Bachettietal21} have recently claimed to find a mass transfer rate $\sim 10^{-6}\msun\,{\rm yr}^{-1}$
in the pulsing ULX (or PULX) system M82 X-2 \citep{Bachettietal14}
by measuring the change of its orbital period over 7 years. 

This claim is based on the standard description of the long--term evolution of a mass--transferring binary system (see e.g. \citealt{King88,Tauris10} and \citealt{Frank02} (Section 4.4), for reviews). 
This uses the equations for conservation of mass and angular momentum for a semidetached binary system, where one star (the mass donor, here denoted as star 2, with mass $M_2$) fills its Roche lobe and transfers mass to the other star (star 1). In the system M82 X-2 star 2 is probably an OB giant, and star 1 is a magnetic neutron star.

\section{Binary Evolution}
Standard formulae (e.g. \citealt{Frank02}) for the size of the donor's Roche lobe, together with its mass--radius relation and Kepler's laws, give relations of the form
\begin{equation}
-\frac{\dot M_2}{M_2} \sim \frac{|\dot R_2|}{R_2} \sim 
\frac{|\dot P|}{P} \sim \frac{1}{t_{\rm evol}},
\label{evol}
\end{equation}
between the mass transfer rate $-\dot M_2>0$, the radius change $\dot R_2$, the orbital period change $\dot P$
and the binary evolution timescale $t_{\rm evol}$. This is given by the stellar evolution of the donor (as is likely in M82 X-2, where this star is probably an OB giant like the progenitor of Cyg X-2, cf \citealt{King99}) or systemic loss of orbital angular momentum. 
Both $\dot R_2$ and $\dot P$ can have either sign depending on what process is driving the evolution, and also the binary mass ratio.

The physical quantity of most interest in studying any interacting binary is the mass transfer rate $-\dot M_2$, but this is frequently hard to measure directly. Then the relations (\ref{evol}) make it tempting instead simply to measure the period evolution $|\dot P|/P$, and deduce $-\dot M_2$ by
using an estimate of $M_2$, and this is what Bachetti et al. (2021) attempt. 

But the
vital point missed by a long line of authors, of whom Bachetti et al. (2021) are the latest, is that these equations hold  
%
\noindent
{\it only when averaged 
 on timescales significantly longer than the time 
 \begin{equation}
 t_{\rm H} = \frac{H}{|\dot R_2|}
 \end{equation}
 for the Roche lobe radius to move one density scaleheight $H$ through the donor star}. 

This requirement comes about because a star does not have a sharp edge ($H = 0$) so cannot be arbitrarily sensitive to very slow secular changes caused by stellar evolution or orbital angular momentum losses. Moreover in deriving (\ref{evol}) it is generally assumed (usually tacitly) that the star is in hydrostatic equilibrium throughout, has no surface magnetic fields, does not lose mass and angular momentum through stellar winds, is not irradiated by the accretor, and is always kept in complete synchronous rotation by tides. 

It is likely that none of these assumptions holds in detail -- for example
the hydrostatic assumption is clearly incorrect precisely where it is needed, near the inner Lagrange point $L_1$, as the star is losing mass at about its local sound speed there. 
So directly observed period changes are actually dominated by large but short--term effects, such as variations in the stellar wind mass loss, oscillations of the stellar envelope, mass currents in the donor driven by irradiation, or magnetic cycles causing slight radius changes. Most of these have no effect on the mass transfer rate $-\dot M_2$. They may be oscillatory and average out, but in all cases are  eventually dwarfed by the systemic binary evolution on timescales $t_{\rm evol} \gg t_{\rm H}$.  The star's gas density $\rho$ near $L_1$ can change exponentially ($\rho \propto e^{\pm t_{\rm evol}/t_{\rm H}}$) on this timescale, the sign depending on whether the binary separation grows or shrinks over time).
This supplies the `sharp edge' forcing $\langle M_2\rangle, \langle R_2\rangle, \langle P\rangle $ and their time derivatives {\it averaged on timescales $\gg t_{\rm H}$} to obey the relations (\ref{evol}).

This point has been made repeatedly in the literature, going back at least to Pringle (1975); see also e.g. Ritter (1988) and Frank et al., (2002, Section 4.4). As an indication of the difficulties involved here, it is common for observed
short--term period changes even to have the wrong sign for the claimed long--term evolution (cf Pringle 1975)\footnote{The orbital periods of cataclysmic variables (CVs) are easy to find as they are only a few hours. Several hundred are now known, but it is well understood in that field that one cannot safely deduce mass transfer rates from them.}.

\section{The ULX M82 X-2}
For an OB giant we have 
\begin{equation}
H \simeq \frac{kTR_2^2}{GM_2\mu m_H} \simeq 5\times 10^7T_4r_2^2m_2^{-1}\,{\rm cm},
\end{equation}
where $M_2 = m_2\msun, R_2 = r_2\rsun$ are the donor mass and radius and $T_4$ its effective temperature in units of $10^4$~K. In fact $H$ is somewhat larger than this near the inner Lagrange point, where the mass transfer takes place. 

Then to find orbital period changes consistent with
the claimed mass transfer rate in M82 X-2, one would need to observe the for a time
\begin{equation}
t \gg t_{\rm H} = \frac{H}{\dot R_2} = \frac{H}{R_2}\frac{R_2}{\dot R_2} = \frac{H}{R_2}t_{\rm evol} 
\gtrsim \frac{10^3T_4r_2}{m_2}\,{\rm yr} >1000\,{\rm yr}
\end{equation}
(where we have adopted \citet{Bachettietal21}'s value 
$R_2/|\dot R_2|\sim 10^6\,{\rm yr}^{-1}$)
rather than 7 yr. For a smaller and probably more realistic binary evolution rate, $t_{\rm evol}$ is even longer, so the required observation time is still more protracted.

\section{Testing the Method}

\citep{Bachettietal21} do not give details of any test of their 
method by using publicly available data on orbital period changes in
mass--transferring binaries.
Three well--studied X--ray transients (Nova Muscae 1991, XTEJ1118+480, A0620--00), all currently in the faint quiescent state, have significant orbital period changes ($-6.56 \times10^{-10}$, $-6.01 \times10^{-11}$, $-1.9 \times10^{-11}$ss$^{-1}$, respectively; \citealt{Gonzales17}), and
so offer a clear test.

Following the argument of \citet{Bachettietal21} these data lead (cf eqn \ref{evol}) to mass transfer rates in the range
\begin{equation}
-\dot M_2 \simeq \frac{|\dot P|}{P}M_2 \simeq 8.5 \times 10^{-9} - 4.8 \times 10^{-7}\, \rm M_{\odot}yr^{-1}.
\end{equation}
These are significantly above the values required
to make the accretion disc in all three systems steady, in the high state where hydrogen is ionized, rather than quiescent and occasionally producing transient outbursts. So they should have bright persistent luminosities $5 \times 10^{37}$ -- $3 \times 10^{39}\, {\rm erg\,s^{-1}}$ (Nova Muscae 1991 would be a ULX!). Yet observation  show that all three systems remain extremely faint\footnote{They are transferring mass at much lower rates \citep{Coriat12}, filling the quiescent accretion disc before its next outburst.}.


\section{Conclusion}

As we have reiterated above, even a millenium is far too short a time to determine the mass transfer rate from orbital period changes in any stellar--mass binary, and M82 X-2 is no an exception. ``... an accurate estimate of the rate of mass transfer cannot be deduced from a change of binary period'' \citep{Pringle75}.  Clear tests of the method of 
\citet{Bachettietal21} using data on period changes in soft X--ray transients give extremely discouraging results. We conclude that
mass transfer value claimed by \citet{Bachettietal21} is physically meaningless as it is deduced from period noise. 

\citet{Bachettietal21} draw far--reaching conclusions from this claim,
in particular that 
all PULXs have a unique property distinguishing them from \textsl{all} \citep{King19} other binary systems, namely a neutron star with  a magnetar--strength magnetic field. But some PULXs are actually ordinary
Be/X--ray binaries when not in particularly bright outbursts, and there is no evidence for such fields in Be/X--ray binaries.

\citet{King19} have shown instead that the properties of 
M82 X-2 
are consistent 
with a neutron star
magnetic field $10^{10} - 10^{11}$~G, a mass transfer rate $\sim 50$ times the Eddington value, and beaming factor $b \sim 0.03$ (see also \citealt{King22}). 
This agrees with the deduction that observed ULXs are ordinary high--mass X--ray binaries in a special stage of their binary evolution \citep{King01,Wiktor17}.


\section*{Acknowledgments}

JPL was supported in part by a grant from the French Space Agency CNES.

{}


\begin{thebibliography}{99}

\bibitem[\protect\citeauthoryear{Bachetti et al.}{2014}]{Bachettietal14} Bachetti M., et al., 2014, Nature, 514, 202  

\bibitem[\protect\citeauthoryear{Bachetti et al.}{2021}]{Bachettietal21} Bachetti M., Heida M., Maccarone T., Huppenkothen D., Israel G.~L., Barret D., Brightman M., Brumback., M., Earnshaw, H.~P., Forster, K., F\"urst, F., Grefenstette, B.~W., Harrison, F.~A., Jaodand, A.~D., Madsen, K.~K., Middleton, M., Pike, S.~N., Pilia, M., Poutanen, J., Stern, D., Tomsick, J.~A., Walton, D.~J., Webb, N., Wilms, J. 2021, arXiv:2112.00339

\bibitem[\protect\citeauthoryear{Coriat, Fender, \& Dubus}{2012}]{Coriat12} Coriat M., Fender R.~P., Dubus G., 2012, MNRAS, 424, 1991

\bibitem[\protect\citeauthoryear{Gonz{\'a}lez Hern{\'a}ndez et al.}{2017}]{Gonzales17} Gonz{\'a}lez Hern{\'a}ndez J.~I., Su{\'a}rez--Andr{\'e}s L., Rebolo R., Casares J., 2017, MNRAS, 465, L15

\bibitem[\protect\citeauthoryear{King}{1988}]{King88} King A.~R., 1988, QJRAS, 29, 1

\bibitem[\protect\citeauthoryear{King et al.}{2001}]{King01} King A.~R., Davies M.~B., Ward M.~J., Fabbiano G., Elvis M., 2001, ApJL, 552, L109. doi:10.1086/320343

\bibitem[\protect\citeauthoryear{King \& Ritter}{1999}]{King99} King A.~R., Ritter H., 1999, MNRAS, 309, 253

\bibitem[\protect\citeauthoryear{King \& Lasota}{2019}]{King19} King A., Lasota J.-P., 2019, MNRAS, 485, 3588

\bibitem[\protect\citeauthoryear{King, Lasota \& Middleton}{2022}]{King22} King A., Lasota J.-P., Middleton, M., 2022, NAR, submitted

\bibitem[\protect\citeauthoryear{Frank, King, \& Raine}{2002}]{Frank02} Frank J., King A., Raine D.~J., 2002, Accretion Power in Astrophysics, CUP

\bibitem[\protect\citeauthoryear{Pringle}{1975}]{Pringle75} Pringle J.E., 1975, MNRAS, 170, 633

\bibitem[\protect\citeauthoryear{Ritter}{1988}]{Ritter88} Ritter H., 1988, A\&A, 202, 93

\bibitem[\protect\citeauthoryear{Tauris \& van den Heuvel}{2010}]{Tauris10} Tauris T.~M., van den Heuvel E.~P.~J., 2010,  ``Formation and evolution of compact stellar X-ray sources", in Compact Stellar X-ray Sources, 623

\bibitem[\protect\citeauthoryear{Wiktorowicz et al.}{2017}]{Wiktor17} Wiktorowicz G., Sobolewska M., Lasota J.-P., Belczynski K., 2017, ApJ, 846, 17


\end{thebibliography}
\end{document}